\documentclass{edmarticle}
\usepackage{graphicx} 
\usepackage{caption} 
\usepackage{algorithm}
\usepackage{algorithmic}
\usepackage{newfloat}
\usepackage{listings}
\usepackage[shortlabels]{enumitem}
\usepackage{calc}
\usepackage{geometry}
\usepackage{float}

\begin{document}

\title{Examining the Influence of Varied Levels of Domain Knowledge Base Inclusion in GPT-based Intelligent Tutors}

\numberofauthors{1}
\author{
\alignauthor
Blake Castleman, Mehmet Kerem Turkcan\textsuperscript{\rm 1}\\
       \affaddr{\textsuperscript{\rm 1}Department of Electrical Engineering, Columbia University, New York, NY\\}
       \email{\{bc3029,mkt2126\}@columbia.edu}
}

\maketitle

\begin{abstract}
Recent advancements in large language models (LLMs) have facilitated the development of chatbots with sophisticated conversational capabilities. However, LLMs exhibit frequent inaccurate responses to queries, hindering applications in educational settings. In this paper, we investigate the effectiveness of integrating a knowledge base (KB) with LLM intelligent tutoring systems (ITSs) to increase response reliability. To achieve this, we design a scaleable KB that affords educational supervisors seamless integration of lesson curricula, which is automatically processed by the ITS. We then detail a preliminary study, where student participants were presented with questions about the artificial intelligence curriculum to respond to. GPT-4 ITSs with varying hierarchies of KB access and human domain experts then assessed these responses. Lastly, students cross-examined the ITSs' responses to the domain experts' and ranked their various pedagogical abilities. Results suggest that, although these ITSs still demonstrate a lower accuracy compared to domain experts, the accuracy of the ITSs increases when access to a KB is granted. We also observe that the ITSs with KB access exhibit better pedagogical abilities to speak like a teacher and understand students than those of domain experts, while their ability to help students remains lagging behind domain experts.
\end{abstract}

\keywords{intelligent tutoring system, GPT tutor, knowledge base inclusion, pedagogical abilities}

\section{Introduction}
Asynchronous education has become increasingly more prevalent over the twenty-first century and is now crucial as a means to assist general education. The surge in online learning, however, has resulted in less student-to-student and student-to-teacher interaction, which has deterred academic engagement \cite{10.3389/fpsyg.2021.713057,doi:10.1080/10494820.2021.1984949}. 

The rise of artificial intelligence (AI) and, in particular, large language models (LLMs), have enabled the construction of intelligent tutoring system (ITS) chatbots, which hold promise in mitigating the latter issue by supplementing pedagogical interactions \cite{DWIVEDI2023102642,10125121}. GPT-3 \cite{brown2020language} and GPT-4 \cite{openai2023gpt4} are some of the largest and most versatile LLMs trained to date, with impressive domain knowledge recall and natural language processing (NLP) abilities. This grants students the ability to pose queries for numerous subjects and receive tailored responses. The potential pedagogical impact on distance learners, therefore, is formidable, with some educational institutions and organizations already announcing the integration of GPT models in various programs \cite{khanacademy,harvardcs50}.

LLMs have several shortcomings, however, that discourage their use in educational settings. Primarily, with GPT models, a lack of domain knowledge or analytical reasoning frequently causes fictitious responses to be generated in their overzealous pursuit of a reply \cite{10125121,zhong2023agieval}. As such, this potentially yields large misunderstandings in educational settings. As a result, LLM ITSs are controversial for present-day classroom application as they have not yet exhibited pedagogical abilities equal to or exceeding that of human teachers \cite{kohnke,tack2022ai}. To develop optimal LLM ITSs, it is crucial to keep LLM responses in alignment with educators' lesson goals and to prevent them from presenting artificial information to students.

This study is the first to examine the efficacy of integrating a knowledge base (KB) with LLM ITSs to achieve these goals. The direct installment of a KB with educators' subject matter may bring GPT to reconsider previously disregarded knowledge, mitigating these issues. Consequently, we aim to explore how integrating KBs may {\it enhance the response accuracy} of LLM ITSs, as well as how their {\it pedagogical abilities may improve} through such integration.

In this paper, we aim to address these open questions by creating GPT-4 ITSs with various KB access levels for accuracy and pedagogical ability comparison. A practice mode was developed for these ITSs in which they assess students' answers to questions about related lesson subject material. The tutors were granted the ability to use any information provided by the supervisor through their KB in the practice mode, given the various information constraints each ITS was assigned. Information constraints were abstracted into a hierarchy of three levels: the lesson's subject (no KB access), a topic description (partial KB access), and the overall lecture material (full KB access).

To evaluate and quantify the effectiveness of each information level in the KB, we populated the LLM ITS with the AI curriculum and designed a three-stage evaluation pipeline to observe the effect of KB integration. First, we designated student participants with AI proficiency to respond to various AI questions with diverse accuracy. Next, human domain expert participants and ITSs assessed the students' responses and addressed each. Lastly, the students were allowed to observe the assessments and rated the pedagogical quality of each response. This allows us to observe the practicality of each knowledge hierarchy level in GPT-4 ITS implementations in contrast to human domain experts. Our code and data are open-sourced below for further research.\footnote{https://github.com/b-castleman/llm-tutor-knowledge-base}.

\section{Related Work}

\subsection{Chatbot Tutor Pedagogical Abilities}

Chatbot tutoring has been applied for the past two decades \cite{Nye,10.3389/frai.2021.654924} and recent research has investigated potential LLM frameworks. Demszky et al. (2021) \cite{demszky-etal-2021-measuring} observed that an ITS based on BERT \cite{devlin-etal-2019-bert}, trained on 2246 student-teacher exchanges, had remarkable conversational uptake to student queries. Tack and Piech (2022) \cite{tack2022ai} built upon Demszky et al. (2021) \cite{demszky-etal-2021-measuring} by discovering that although LLMs such as Blender \cite{roller-etal-2021-recipes} and GPT-3 have considerable conversational uptake, core pedagogical abilities of these LLMs severely lag behind that of human teachers. They discovered these LLMs demonstrate insufficient capability in their pedagogical abilities to speak like a teacher, understand students, and help students. We aim to investigate if the substandard pedagogy measurements Tack and Piech (2022) \cite{tack2022ai} have found can be improved upon by expanding the knowledge these LLMs have to work within.

\subsection{Chatbot Tutor Knowledge Base Incorporation}

Using KBs for chatbot tutors to retrieve information is not a novel topic. An instance of such is AsasaraBot \cite{app12073239}, an educational AI chatbot designed to teach high school students cultural content in foreign languages, which searches its own KB (or externally on the Web) for the most appropriate response and processes it within its dialogue manager.

BookBuddy \cite{10.1145/3330430.3333643}, a virtual reading companion designed to teach English as a second language, uses a scaleable knowledge base of 20 books to choose from (depending on the child's interests). Trained on the Stanford Question Answering Dataset (SQuAD) \cite{rajpurkar-etal-2016-squad}, BookBuddy's generative model handles character and plot questions, while its rule-based algorithms produce vocabulary and arithmetic questions.

QuizBot \cite{10.1145/3290605.3300587}, a dialogue-based agent that helps students learn factual knowledge, uses a system question pool populated by a supervisor's question-answer pairs that pertain to the lesson subject. It was designed to binary grade student answers via cosine similarity thresholding and was observed to rarely misclassify answers. In their studies, they evaluated the ability of the agent to generalize over three different subjects and data suggested QuizBot's effectiveness to surpass that of flashcard techniques.

In this paper, we seek to quantify the efficacy of a GPT ITS implemented in conjunction with a knowledge base. Furthermore, we aim to observe the variations in our tutor's response quality from differing quantities of inputted supervisor information presented by the knowledge base through our access hierarchy framework. To our knowledge, there are currently no published chatbot tutors that have researched the effectiveness of teaching artificial intelligence course curricula or that have quantified the effect of disparate KB hierarchy levels on response quality.

\section{Framework}

\subsection{Prompting Structure}

We use GPT-4 to create three separate LLM ITS instances. It is prompted to respond ``as a teacher", being both accurate and enthusiastic. The tutor operates in a practice mode, where students are quizzed by the tutor about the material to reinforce and consolidate their understanding of what they have learned. After receiving a student's answer to a particular question, the tutor must respond appropriately depending on the student's answer choice(s) and reasoning. After the student responds to the question, GPT-4 ranks the responses on an integer scale from 1-5 and stores this value internally. Then, it is asked to qualitatively assess and address the student's answer given the previous rating, describing what is understood well by their answer and what may use improvement. This internal rating technique is inspired by Park et al. (2023) \cite{park2023generative}, who used situational GPT-4 ratings to simulate autonomous agent behavior. 

\subsection{Knowledge Base}

\begin{figure}
\begin{center}
\includegraphics[width=0.95\linewidth]{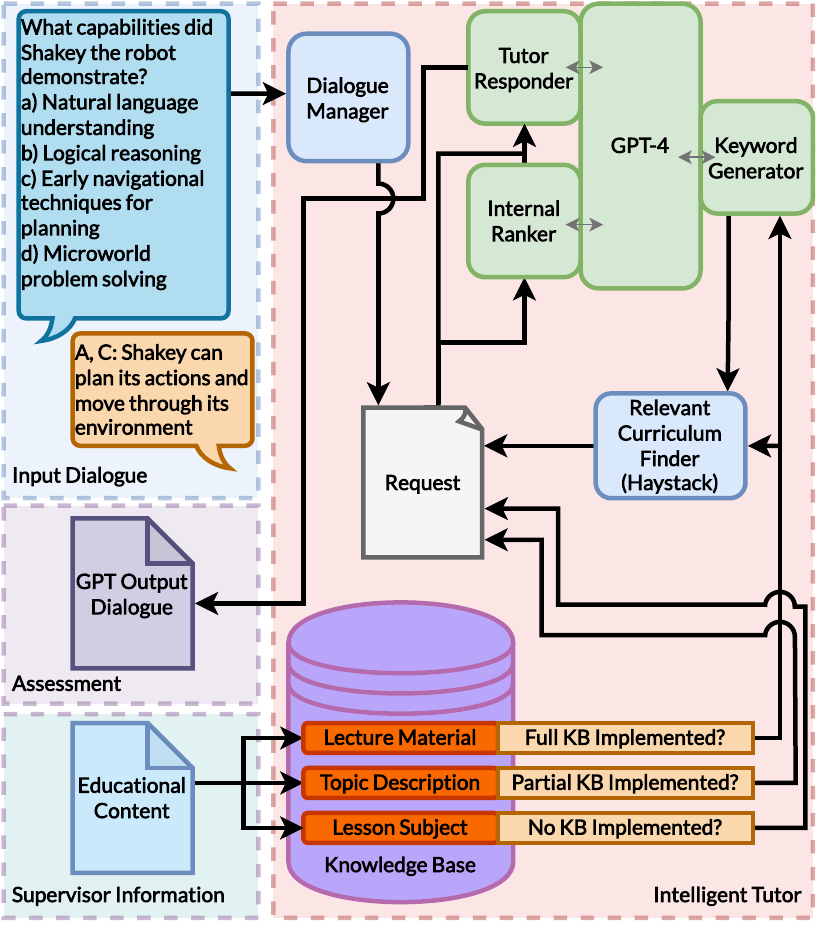}
\caption{Schematic of the ITS architecture implemented. }
 \label{final-schematic} 
\end{center}
\end{figure}

The KB is designed for an educator to efficiently input their lesson material into JSON files for information processing at disparate knowledge hierarchies. We will now describe how each hierarchy level of information provided interacts separately with the system. Figure \ref{final-schematic} illustrates the following described architecture to help understand the information processing pathways implemented. 

\subsubsection{No KB Access: Lesson Subject}

The base-level hierarchy, hereon referred to as having no KB access, is told only what the lesson subject is (i.e. introduction to artificial intelligence, basic derivatives, prokaryotic cells). As a result, any ITS that uses only this KB level is only using GPT-4 training with minimal supervisor influence to generate responses. The lesson subject is given to the tutor when defining the initial system prompt and is repeated when internally rating any answer.

\subsubsection{Partial KB Access: Topic Description}
The topic description is a short definition of what the given topic is about, consisting of no more than a few sentences. Any tutor using KB levels up to this extent is referred to as having partial KB access as it can observe high-level information on the lesson subject without being presented with any intricacies. The topic description is used when defining the initial system prompt and is given to the response ranker.

\subsubsection{Full KB Access: Lecture Material Inclusion}
The lecture material is defined as the text copy of a lecture's transcript. Any tutor using KB levels up to this extent is referred to as having full KB access as the full extent of the supervisor's lecture material is readily available to the tutor. The lecture material is incorporated into the ITS's implementation as follows:

\begin{enumerate}
    \item \textbf{Subtopic Separation}

    Due to the number of tokens that lecture transcripts often contain, we divide lecture content into subtopics. In the implemented design, the supervisor manually performs this and can create as many subtopics as they deem fit, given they restrict the number of tokens in individual subtopics to prevent prompting overflow.

    \item \textbf{Keyword Generation}

    To make the subtopics more searchable, we automatically generate lists of keywords that pertain to each subtopic. We performed this operation using GPT-4, querying it to create a set of keywords that best represents the subtopic and data included. We then concatenate it to the subtopic for processing.

    \item \textbf{Haystack Question-Answering}
    We utilize Haystack by Deepset \cite{haystack}, an open-source NLP framework that offers the ability to yield complex question-answering capabilities using transformer models and LLMs. This framework allows us to utilize RoBERTa \cite{liu2019roberta}, a state-of-the-art BERT LLM trained on the SQuAD dataset, to automatically find the corresponding subtopics within the lecture material that is relevant to the student's response. Each subtopic at this stage is now composed of the associated lecture material, chosen by the supervisor, and the keywords, generated by GPT-4. Having both pieces of data included in the search assists in making the relevant subtopics readily discoverable. We use the Haystack approach in parsing the KB in order for the ITS to be a) scaleable (unlikely to near GPT-4's token limit) and b) low-cost (computationally and monetarily). After finding the appropriate subtopic(s), they are concatenated onto the original lecture material for the internal ranking of the student's answer and for the qualitative assessment of the student's answer.
\end{enumerate}

\section{Study Design}

\subsection{Artificial Intelligence Course Content}

To evaluate the effectiveness of each information level in the KB, we populated the LLM ITS with the boot camp curriculum from Aiphabet: a secondary-school informal learning platform for teaching artificial intelligence \cite{aiphabet}. It was founded on the course content for Columbia University's undergraduate artificial intelligence course and the curriculum is curated by many of the university's professors and graduate students. Specifically, we chose to use their video lecture titled ``The History of AI" as our lesson subject as it was deemed to be the least subjective of the current prerelease materials. The transcript was then divided into four subtopics according to the lecture's described central eras of AI's history.

\subsection{Model Instantiations}

To obtain the necessary GPT-4 ITSs, we generated three identical tutors, varying only by the KB permissions they are allocated. The first model was only granted access to the lesson subject (no KB access), the second model was granted access to the lesson subject and the topic description (partial KB access), and the third model was granted access to the lesson subject, topic description, and lecture material (full KB access). This allowed the LLM response qualities to be observed with respect to the addition of each knowledge level.

\subsection{Evaluation Pipeline}

We designed a three-stage pipeline to accomplish the evaluation of the three ITSs. Note that all data was collected asynchronously, being delivered and received over writable documents. Appendix Figure \ref{pipeline-table} portrays a sample of a participant's final ratings of two assessor responses with the entire evaluation pipeline's stages intertwined. We will now discuss how each stage was conducted:

\subsubsection{Stage \#1: Student Question-Answering}

For the first stage, we recruited four university students who had completed a college-level AI course for the evaluation. Participants' AI knowledge varied from undergraduate computer science students casually interested in AI to Ph.D. candidates who specialized in it, all of whom volunteered and agreed to the anonymized release of their collected data.

First, each participant was asked to watch the aforementioned video lecture with relevant course material. They were then given a total of nine questions (creation detailed in Appendix \ref{evaluation-question-creation}) with an approximately uniform amount from each subtopic being randomly and uniquely picked for each student. Every participant was then asked to respond to each question presented to them with varying levels of accuracy. This was realized by asking participants to answer a third of their questions with full accuracy, a third of their questions with partial accuracy or partial inaccuracy, and a third of their questions very inaccurately. We prompted participants for a variety of accuracy levels in order to ensure a similar corpus of response qualities was achieved, which was necessary to properly assess whether or not the KB could react proficiently to students' answer slips and guesses. No other restrictions were imposed on answers. No time limit was imposed for participants' answers.

\subsubsection{Stage \#2: Domain Expert and Intelligent Tutor Addressal}

In stage two, we recruited two professional AI curriculum domain expert volunteers who agreed to the anonymized release of their collected data. Both domain experts were professionals in AI education with multi-year experience in curating and teaching AI course curricula for secondary and post-secondary students. Additionally, both domain experts agreed they had the necessary expertise to properly assess and address potential student responses to the lesson material and provided questions. Before beginning the evaluation, we secured explicit confirmation from the domain experts on their agreement to the correctness of the video lectures that would be provided to students. This validation process ensured a robust foundation for the subsequent phases of our research, reinforcing the reliability of data collected. We utilized these two domain experts as well as the three ITSs (all five henceforth collectively denoted as ``assessors") to assess and respond to students' answers. Specifics on assessors' instructions can be found in Appendix \ref{assessor-prompts-limitations}.

\subsubsection{Stage \#3: Ranking of Assessor's Responses}

For the third stage, the student participants from Stage \#1 evaluated which assessor's response (after processing as per Appendix \ref{assessment-processing}) was of the best quality for each question-answer pair provided. The assessments were anonymized to prevent biases, and the corresponding question-answer pairs were presented with each assessment as contextual background. Due to an overwhelming amount of data to present to students, we uniformly and randomly selected only one domain expert's answer to present to the students for comparison against the ITSs' responses (Appendix \ref{domain-expert-randomized-choice}). 

The triples (question-answer-assessments) given to each participant consisted of those containing the answers they provided to questions in Stage \#1 along with the triples to an additional, random participant's question-answer pairs. This allowed for each assessor's response to have two independent evaluator's rankings and mitigated personal biases.

As every question-answer pair had four separate replies (three ITSs and one randomized domain expert) to rank, we formatted this evaluation as a pairwise comparison between each ITS versus the domain expert. This allowed us to observe how students' assessor preferences changed with KB differences. We utilized the same criteria implemented in past literature and asked participants which response is more likely to be said like a teacher, understanding of a student, and helpful to a student \cite{tack2022ai}, which builds on key pedagogical requirements set forth for AI models being applied as AI teachers \cite{Bommasani2021FoundationModels}. For each pedagogical ability criterion, we asked participants to indicate whether either assessment was of better quality or if they were too similar to differentiate.

Furthermore, a considerable amount of assessments suggested inaccurate answers as truthful information. To ensure participants' familiarity with which choices were valid and rational, a ``Factual Information" section was included for each question. This section gave a brief delineation, based on rigorous research, for the answer choices that most accurately addressed the given question. The assessors' replies were never directly mentioned in this section to reduce biases.

\section{Results}

\subsection{Stage \#2 Results: Agreement and Accuracy}

We cross-compare the domain experts' assessments against one another to find an agreement level of 63\% for what each domain expert believed the correct answer choices were whereas the ITSs only agreed 43\% of the time on what they believed the correct answer choices were. Some of the answer choices were ambiguous depending on the specific context that each assessor interpreted (e.g. the potential impact of a technology to have indirectly influenced events), meaning assessor groups would not always be in agreement in their responses. In calculating these agreement level percentages, we choose for agreements to be strict and for any assessor response deviance to not count as agreement. 

As a result of this inflexible and conservative definition, we also calculate the overall accuracy level for all assessors, taking into account these ambiguous situations. In this definition for accuracy, we declare that ambiguous situations are acceptable so long as no evidence renders the response incorrect or unreliable beyond a reasonable doubt. With this definition, we found that domain experts demonstrated a combined accuracy rate of approximately 87\% while ITSs exhibited a combined accuracy rate of only 70\%. In particular, the no KB, partial KB access, and full KB access ITSs had accuracy scores of 67\%, 77\%, and 74\%, respectively. Information on ITS token usage is available in Appendix \ref{stage2-token-usage}.

\subsection{Stage \#3 Results: Pedagogical Abilities}

\begin{table*}
\begin{center}
\begin{tabular}{ |c|c|c|c|c|  }
 \hline
  Intelligent Tutor & Pedagogical Ability & GPT-4 Preference & Domain Expert Preference & No Clear Preference\\
 \hline
    & Talking like a teacher    & 27.14\% &   \textbf{50.00\%} & 22.86\%\\
 No Knowledge Base &   Understanding the student  & 31.43\%   &\textbf{38.57\%} & 30.00\%\\
  &   Helpful to the student  & 25.71\%   &\textbf{57.14\%} & 17.14\%\\
 \hline
  &   Talking like a teacher  & \textbf{44.29\%}   &38.57\% & 17.14\%\\
 Partial Knowledge Base &   Understanding the student  & \textbf{41.43\%}   &35.71\% & 22.86\%\\
  &   Helpful to the student  & 38.57\%   &\textbf{51.43\%} & 10.00\%\\
 \hline
  &   Talking like a teacher  & \textbf{38.57\%}   &31.43\% & 30.00\%\\
 Full Knowledge Base &   Understanding the student  & \textbf{41.43\%}   &31.43\% & 27.14\%\\
  &   Helpful to the student  & 34.29\%   &\textbf{45.71\%} & 20.00\%\\
 \hline
\end{tabular}
\caption{Preference rates for different tutors based on evaluations of four participant annotators for each pedagogical ability.}
\label{pedagogy-table}
\end{center}
\end{table*}

We examine the preliminary results by calculating the percentage of replies classified as a particular pedagogical ability for each ITS (Appendix \ref{stage-3-data-processing}). Table \ref{pedagogy-table} exhibits the results of this computation and Appendix Figure \ref{pedagogy-plot} showcases a bar graph of the rankings to illustrate the visual change in reply selection quantity with KB additions.

We observe, for the ITS with no KB access, that the domain experts attained a relative majority for all pedagogical abilities, particularly in the dimension of being helpful to students. This agrees with Tack and Piech (2022) \cite{tack2022ai}, who found similar trends in pedagogy for GPT-3 AI teachers responding to student dialogues. Though their LLM (GPT-3) and our own (GPT-4) are disparate, as well as the domain of education, the exhibition of their similar pedagogical qualities underscores the consistency of these didactic models across different iterations and applications.

Upon integration with a KB, the ITSs' pedagogical abilities exhibited increases in preference for GPT-4 responses. The partial KB ITS achieved 17.15\%, 10.00\%, and 12.86\% preference increases for talking like a teacher, understanding the student, and being helpful to the student respectively. Meanwhile, the full KB ITS achieved 11.43\%, 10.00\%, and 8.58\% preference increases, respectively. This suggests that having both partial KB access and full KB access increases the ITSs' pedagogical dimensions.

Notably, the reply percentage distributions for talking like a teacher and understanding the student realigned in favor of both ITSs with KB access. No ITS's ability to be helpful to the student overcame the domain experts', despite the other two abilities accomplishing this task.

The partial KB ITS was particularly preferred for talking like a teacher and tied with the full KB ITS for understanding the student. However, the domain expert reply preferences were consistently lower for the full KB ITS as opposed to the partial KB ITS for all three pedagogical abilities.

\section{Discussion and Future Work}
Our study faced a core limitation in the evaluation due to a shortage of AI curriculum experts with adequate knowledge and time allocations. Consequently, we had to reduce the number of student participants and prompt students for a variety of response accuracy to ensure we received assessor responses on incorrect answers. This approach purposefully ensured a generation of sufficient inaccurate student answers for assessors to evaluate. However, trying to mimic naturally inaccurate student answers (i.e. slips and guesses) with artificial inaccuracies could have potentially introduced experimental bias. Ideally, in the scenario of more personnel resources becoming available, future studies of KB integration should have more AI curriculum domain experts and student participants to resolve the aforementioned issues.

We also chose to include participants who have completed courses in AI instead of participants engaged in an AI course for the first time. This afforded participants a preferable capability to contrast the pedagogical abilities presented in Stage \#3, where seemingly correct arguments from assessor responses could mislead students who lack prior, formal education in AI. Moreover, this decision raised the likelihood that the inaccurate student answers provided in Stage \#1 were plausible and well-articulated.

Using ``check all that apply" questions were used to make assessors required to evaluate multiple components of every question in Stage \#2. As a result, it should be noted that other question types may produce different results. Dissimilar curricula may also produce different results as question complexity and response variation depend on them.

Lastly, though KBs external to LLMs possess multifaceted benefits (e.g. ease of updates, direct prompting) fine-tuning these models on educational subject matter may prove incredibly resourceful for accuracy and pedagogical ability gains. We hope to investigate this possibility in future research.

\section{Conclusion}

Our study is the first to delve into the extent to which an integrated KB can enhance the pedagogical abilities of LLM chatbot tutors. This technique seeks to mitigate inaccuracies and irrelevant responses from LLMs, thereby potentially increasing pedagogical abilities. We gather preliminary data on such systems by exploring, at varying KB hierarchies, how effective and accurate the ITSs are compared to domain experts. An evaluation was conducted on AI curriculum utilizing both ITSs (with varying KB access constraints) and domain experts to assess student answers to various ``check all that apply" question types for a given curriculum.

The assessor responses retrieved from our evaluation pipeline indicated that all ITSs had lower agreement and accuracy rates than domain experts. Both ITSs with KB access, however, had notably higher accuracy rates as compared to the ITS without KB access. After quantifying the ITSs' pedagogical abilities, we observe, for the aforementioned curriculum and question types, that the KB integration increases the preference for ITS responses and decreases the preference for domain expert responses. Furthermore, our data suggested that the abilities of ITSs to speak like a teacher and understand students surpassed that of domain experts when KB capabilities were used. The ability to be helpful to students, though increasing with KB inclusion, still failed to exceed that of domain experts.

This comprehensive evaluation highlights the practicality of implementing KBs with chatbot ITS frameworks using LLMs. Though LLM ITSs still have necessary development for application, this methodology spotlights a scaleable technique for instructors to include their course information with chatbot ITSs that results in accuracy and pedagogical ability gains.

\section{Acknowledgments}
We would like to thank the organization Aiphabet for providing their prerelease AI curriculum for our use. We also would like to thank the anonymous domain experts and student participants for their invaluable contributions to the study. 

\bibliographystyle{abbrv}
\bibliography{sigproc}

\appendix

\section{Stage \#1: Question Creation}

\label{evaluation-question-creation}

To create impartial yet relevant questions, we used GPT-4 to generate a set of 36 ``check all that apply" questions. This question type was chosen as it yields an exponential amount of answer possibilities, increasing the likelihood that any of the assessors may evaluate an answer inaccurately. With the expansive question bank, the influence of outliers is minimized and correlations in assessor agreements can be observed.

As this information alone may not have been sufficient for understanding a student's weaknesses from incorrect responses, we prompted participants to explain their answers for every answer given. The question relevancy was equally divided amongst the four subtopics, and GPT-4 was given full KB access to the particular AI subtopic as well as the topic description for generating each question. This best mimicked question creation by various educators, who often have abundant resources (i.e. textbooks, internet) for creating intricate questions.

Before proceeding, we went through all generated questions to ensure they were valid for different levels of difficulty, pedagogically sound, and relevant to the video lecture's material. Five questions were chosen to be regenerated for being either repeats of other questions or for lacking relevance to the given lecture material.

\section{Stage \#2: Assessors' Prompt and Instructions}

\label{assessor-prompts-limitations}

We utilized the two domain experts as well as the three ITSs to assess and respond to students' answers. The five assessors were given the same prompt and were asked to do the following:

\begin{itemize}
    \item Respond to the student's answer given the current question
    \item Explain what about their answer is accurate or inaccurate
    \item Suggest what about the answer can use improvement
    \item Help clear any misconceptions
\end{itemize}

Assessors were limited to a 75-word constraint and had no time limit. The only difference in the prompting for the ITSs opposite to the domain experts was the ITS's access to an internal ranking of the answer.

\section{Evaluation Pipeline Stages}

\begin{figure}[H]
\begin{center}
\begin{tabular}{ |p{7.5cm}|  }
 \hline
  \begin{center} \textit{Question Generation} \end{center}
  \textbf{Question:} What does a neuron output if the sum of its inputs does not exceed the threshold value?  \\

\begin{center}
a) 0 \quad
b) 1 \quad
c) -1 \quad
d) The threshold value
\end{center}\\
\\
\hline
  \begin{center} \textit{Stage \#1} \end{center}
  \textbf{Student Answer:} a - the mcculloch pitts artificial neuron only fires if its input sum exceeds the threshold value, otherwise it remains a zero
  \\
  \\
  \hline
  \begin{center} \textit{Stage \#2} \end{center}
  \textbf{Reply A:} That's correct! Your understanding of the McCulloch-Pitts artificial neuron is accurate. The neuron outputs a 0 if the input sum does not exceed the threshold value. \\
  \\
  \textbf{Reply B:} Great job, that’s correct! In the context of a neuron with a threshold, assuming the McCulloch and Pitts model, both the inputs to the neuron and the response from the neuron are all binary (0 or 1). 1 means the neuron fires (on) and 0 means the neuron doesn’t fire (off). \\
  \\
  \textbf{Factual Information:} A is the only correct answer as artificial neurons (for McCulloch \& Pitts) output a zero for insufficient aggregate values. \\
  \\
  \hline
  \begin{center} \textit{Stage \#3} \end{center}
  Which of these replies is more likely said by a teacher?\\
 \begin{center}   \fbox{\textbf{I cannot tell}} \end{center} \\ 
  In which reply is the teacher understanding the student more? \\
  \begin{center}   \fbox{\textbf{A}}    \end{center} \\
    In which reply is the teacher helping the student more? \\
\begin{center}  \fbox{\textbf{B}} \end{center}\\ 
\hline
\end{tabular}
\caption{A depiction of the evaluation pipeline stages alongside a sample of a student participant's rating for assessor replies.}
\label{pipeline-table}
\end{center}
\end{figure}

\section{Stage \#3: Overwhelming Data for Comparison}

\label{domain-expert-randomized-choice}

For the third stage, we uniformly and randomly selected only one domain expert's answer to present to the students for comparison against the ITSs' responses. We roughly estimated that each pairwise comparison would take one minute to analyze and interpret for meaningful results, meaning that our participant volunteers would likely spend closer to two hours for the third stage evaluations in addition to their earlier contributions in the first stage. Concerns grew that an estimated time requirement exceeding an hour may perhaps induce boredom in the participants and would yield more haphazard data. Therefore, our effort to minimize comparisons attempts to preserve our data's integrity.

\section{Stage \#3: Assessor Response Processing}

\label{assessment-processing}

Before relaying the assessors' replies to the participants, all replies were processed as according to the following list:
\begin{itemize}
    \item Fixed all spelling and grammar mistakes
    \item Removed all sentences of pure affirmation or motivation (i.e. ``Keep up the good work!")
    \item Removed all declarations affirming the student's knowledge level in AI as a whole (i.e. ``Your understanding aligns well with the biological basis of artificial neurons.")
    \item Removed impractical assessment scenarios (i.e. ``How would you change your explanation? [Pause and give time to answer]")
\end{itemize}
Though student participants were never informed that any reply stemmed from GPT, we suspected they may intuitively be able to recognize common GPT-generated sentence structures against human domain expert sentence structures. Thus, these changes were targeted in order to further anonymize the study and conceal which replies were from domain experts or from LLMs. 

\section{Stage \#2 Results: Token Usage}

\label{stage2-token-usage}

We find the computational price for the described architectures of the no KB, partial KB access, and full KB access ITSs to call GPT-4 to be approximately $400$, $600$, and $2000$ input tokens per student answer assessment, respectively. Additionally, all ITSs output approximately $60$ tokens per student answer assessment. This corresponds to total monetary prices of about $\$0.016$, $\$0.022$, and $\$0.064$ per student answer assessment, respectively, as present-day GPT-4 token costs being $\$0.03$ per $1000$ input tokens and $\$0.06$ per $1000$ output tokens.

\section{Stage \#3 Data Processing}

\label{stage-3-data-processing}

We examine pedagogical ability results by calculating the percentage of replies classified as a particular pedagogical ability for each ITS. This allows us to isolate the probability an ITSs' responses will be preferred or unfavored by a student. The calculation is performed by finding the percentage of times each reply option (GPT preference, domain expert preference, no clear preference) was selected within a pedagogical ability category (speaking like a teacher, understanding the student, helpful to the student) for a specific ITS (no KB access, partial KB access, full KB access). Mathematically, this is derived by dividing the total count of selections for a reply option for an ITS-pedagogical ability pair by the total count of all rated replies for that particular ITS. This number is then converted into a percentage for the particular ITS-pedagogical ability so that it can be.

For example, for the ITS with no KB access in the talking like a teacher pedagogical category, GPT-4 preference was chosen $19$ times, and a total of $70$ ratings were given for that ITS in that category. This yields a percentage of $\frac{19}{70} \times 100\% = 27.14\%$

\section{Rankings Graph Visualization}

\begin{figure}[H]
\begin{center}
\includegraphics[width=.98\linewidth]{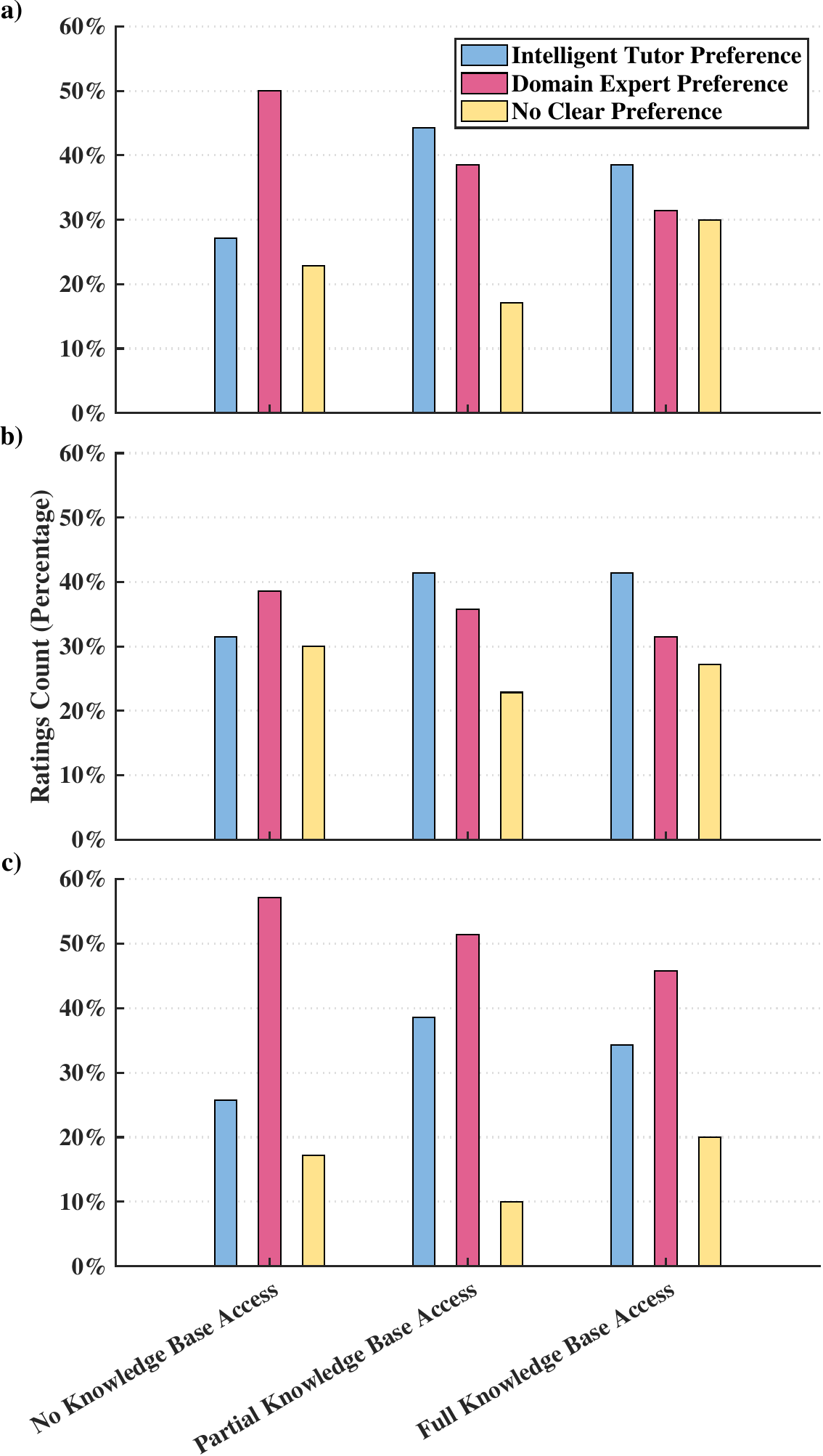}
\caption{Visualization of ITSs' pedagogical ability improvements as knowledge base access increases for a) speaking like a teacher, b) understanding the student, and c) being helpful to the student.}
\label{pedagogy-plot}
\end{center}
\end{figure}

\end{document}